\documentclass[aps,prb,twocolumn,superscriptaddress,showpacs,amsmath,amssymb,longbibliography]{revtex4-1}
\usepackage{graphicx}
\usepackage[breaklinks]{hyperref}
\usepackage[english]{babel}
\usepackage{amsmath}
\usepackage{xcolor}
\usepackage{color}
\definecolor{nred} {RGB}{224,0,0}
\definecolor{nblue} {RGB}{28,130,185}
\definecolor{dgreen} {RGB}{78,138,21}

\usepackage[sort&compress]{natbib}
\begin{document}

\title{Transient and persistent particle subdiffusion in a disordered chain coupled to bosons}
\author{P. Prelov\v sek}
\affiliation{J. Stefan Institute, SI-1000 Ljubljana, Slovenia }
\affiliation{Faculty of Mathematics and Physics, University of Ljubljana, SI-1000 Ljubljana, Slovenia }
\author{J. Bon\v ca}
\affiliation{Faculty of Mathematics and Physics, University of Ljubljana, SI-1000 Ljubljana, Slovenia }
\affiliation{J. Stefan Institute, SI-1000 Ljubljana, Slovenia }
\author{M. Mierzejewski}
\affiliation{Department of Theoretical Physics, Faculty of Fundamental Problems of Technology, Wroc\l aw University of Science and Technology, 
50-370 Wroc\l aw, Poland}

\begin{abstract}
We consider the propagation of a single particle in a random chain, assisted by the coupling
to dispersive bosons. Time evolution treated with rate equations for hopping between localized states reveals
a qualitative difference between dynamics due to noninteracting bosons and hard-core bosons. In the first
case the transient dynamics is subdiffusive, but multi--boson processes allow for long-time normal diffusion, while
hard-core effects suppress multi--boson processes leading to persistent subdiffusive transport, consistent with 
numerical results for a full many-body evolution. In contrast, analogous study for a quasiperiodic  potential reveals a stable 
long-time diffusion.

\end{abstract}

\maketitle

\section{Introduction}
Single particle (SP) localization in a random potential is a well understood phenomenon since the seminal works of 
Anderson \cite{anderson58} and Mott.\cite{mott68} It has become a novel challenge since the proposal of 
many-body localization (MBL) \cite{basko06,oganesyan07}  which would persist in the presence of particle interaction. \cite{pal10,znidaric08,barisic10,gramsch12,serbyn13,lev14,prelovsek217,huse13,deluca13,deluca14,bardarson12,kjall14,
Serbyn13_1,huse14,serbyn141}
As the limiting case for the MBL physics, one can consider a single particle in a disordered system
coupled to bosonic (or other) degrees of freedom.  This problem has a long history related to the phonon-assisted 
variable-range hopping.\cite{mott168} However, there is recently an increasing interest
due to limitations of the validity of this concept \cite{banerjee16} in disordered lattices and due to its relation to MBL 
physics,\cite{lev15,kozarzewski16,serbyn15,khemani15,gornyi05,altman15,rademaker16,chandran15,ros15,eisert15,zakrzewski16,
prelovsek2018,Abanin2017,Abanin2015,Ponte2015,Bordia2017,imbrie2017,obrien16,inglis16,gluza2018,mierzejewski2018,choi16}
in particular because of  
anomalous subdiffusive transport.\cite{bonca17,gopal17,lemut17,bonca18,bordia2017_1,new_karrasch}
In general, the coupling to itinerant (dispersive and non--localized) 
phonon modes leads to the delocalization of the particle. \cite{mott168,banerjee16} This has recently been tested both for a particle 
in  one-dimensional (1D) disordered chain, coupled  to noninteracting bosons (NB),\cite{bonca18} as well as for particle 
dynamics in a $t$-$J$ chain. \cite{bonca17,lemut17} The latter case represents the coupling to $S=1/2$ spins, or 
equivalently hard-core bosons (HCB). On the other hand, localization of bosons modifies the  variable-range hopping,\cite{banerjee16} 
being also the case for
coupling to nondispersive phonons \cite{bonca18} or to localized spin subsystem.\cite{lemut17}

Although the evidence above shows that the SP localization is unstable  against the coupling to dispersive bosons
or, in general,  to a heat bath,\cite{nandkishore14,nandkishore15,huse15} the particle dynamics can still be anomalous. Namely, 
there are examples and regimes where the transport is subdiffusive, i.e., the d.c. mobility vanishes since the spread at
long times behaves as $\sigma^2(t) \propto t^\gamma$ with $0<\gamma<1$.  It has been shown, e.g.,  that a SP subject to local random
noise\cite{gopal17} can exhibit a long transient subdiffusion before turning into a normal diffusion. Similar transport has
been found also for spins on a Hubbard chain with a potential disorder\cite{prelovsek16}, originating in a singular distribution 
of effective exchange couplings.\cite{kozarzewski18} Such a Griffiths--type mechanism for subdiffusion has 
been invoked also for the ergodic side 
of the 1D Heisenberg model with random magnetic fields,\cite{agarwal15,luitz16,gopal16,agarwal16,znidaric16,luitz116} although 
some results indicate that this might be a transient feature to normal diffusion.\cite{steinigeweg16,barisic16,prelovsek116}

In this paper we consider the propagation of a SP in a random chain, coupled to dispersive bosons, which can be either NB
or HCB, whereby the latter case simulates coupling to spins.  We analyze the dynamics in terms of the rate equations for the particle 
hopping between the Anderson eigenstates. The transition rates are evaluated via the Fermi-golden-rule (FGR), but taking into 
account the actual Anderson eigenstates and multi-boson processes. Our main result concerns the essential difference between 
NB and HCB models. In the first case, the long-time limit is shown to be diffusive 
with $\sigma^2 \propto t$.  Nevertheless  the evolution is subdiffusive within the initial time-interval $t<t^*$, where  $t^*$ may 
be very  large depending on disorder and temperature $T$. On the other hand, the  HCB
reveal persistent $\gamma < 1$ for not too weak disorder. Subdiffusion is well resolved also in the numerical
evolution of the whole many-body quantum system. Still, we find that the propagation depends on the details of  potential 
distribution. In contrast to the HCB case with random uncorrelated potentials, the quasi-periodic potential (as relevant for actual MBL 
experiments on cold fermions\cite{schreiber15,bordia16,luschen17}) induces the diffusion in the long-time limit.

\section{Model}
We study the Anderson model for a SP moving in a 1D random potential and 
coupled to boson degrees, 
\begin{eqnarray}
H &=& - t_h \sum_i (c^\dagger_{i+1}  c_i + {\mathrm h.c.}) + \sum_i h_i n_i  + g \sum_i n_i 
( a^\dagger_i + a_i ) \nonumber \\
&+& \omega_0 \sum_i a_i^\dagger a_i - t_b  \sum_i ( a_{i+1}^\dagger a_i + \mathrm{h.c.}),
\label{pbm} 
\end{eqnarray}   
where $n_i=c^\dagger_i c_i$ is the local particle number. Bosons with $\omega_0 >0$ are dispersive 
due to  hopping, $0<t_b < \omega_0/2$. We consider further on two cases: a) NB with a standard
boson
Hamiltonian $H_b= \sum_q \omega_q a^\dagger_q a_q$ and $\omega_q = \omega_0 - 2 t_b
\cos q$, and b) HCB which have restricted Hilbert space with only two states per site 
(formally $a_i^\dagger a_i^\dagger = a_i a_i =0$). Effectively, HCB
represent a spin $S=1/2$ XY chain (in magnetic field $\omega_0>0$) closely related 
to the low-doping limit  of the disordered $t$-$J$ or  $U\gg t_h$ Hubbard models.\cite{mondaini15,prelovsek16,bonca17} 
In the following we put $t_h=1$ while the potentials are uncorrelated and 
uniform with $-W< h_i <W$. It makes sense to rewrite Eq.~(\ref{pbm}) in 
the Anderson basis, 
\begin{eqnarray}
H&=&H_{\rm SP} + H^\prime +H_b, \nonumber \\
H_{\rm SP}&=&\sum_l \epsilon_l \varphi_l^\dagger \varphi_l ,\quad H' = 
 \sum_{l l' i} \eta_{l l' i} \varphi_{l'}^\dagger  \varphi_l (a^\dagger_i + a_i) , 
\end{eqnarray}
where $ \varphi_l = \sum_i \phi_{li} c_i$ are operators of Anderson localized 
states (with real $\phi_{li}$), and  $\eta_{l l' i}  =  g \phi_{l'i}  \phi_{li}$.

\section{Noninteracting bosons}
\subsection{Transition rates}
 In the case of  NB we proceed  by introducing normal modes,
\begin{equation}
H' = \sum_{l l'} \varphi_{l'}^\dagger  \varphi_l   H'_{ll'}, \qquad  H'_{ll'} = \sum_{ q} (\eta_{l l' q}  a^\dagger_q +  \eta^*_{l l' q}  a_{q}), 
\end{equation}
with $\eta_{l l' q}  =  (g /\sqrt{L}) \sum_{i}  \mathrm{e}^{-iqi} \eta_{l l' i}$. Separating $H'$ into the diagonal  part $H'_d$ 
with $l=l'$ and the off--diagonal one, we  transform out $H'_d$  via  standard canonical transformation, 
\begin{equation}
\tilde H= \mathrm{e}^S H \mathrm{e}^{-S}, \quad S = \sum_l n_l A_l, \quad  A_l= \sum_k [\zeta_{l k} a^\dagger_k - \zeta^*_{l k} a_k],
\end{equation}
with $\zeta_{l k} = \eta_{ll k}/\omega_k$. This leads to transformed $\tilde H'$ 
relevant for transitions
\begin{equation}
\tilde H^\prime_{ll'}=  \sum_{q}  \mathrm{e}^{-A_{ll'}} (\eta_{l l' q} a^\dagger_q + \eta^*_{l l' q} a_q ) ,
\label{hbll}
\end{equation}
with $A_{ll'}=A_{l'}-A_l$. Assuming slow transition rates $\Gamma_{ll'}$ between states 
with $\Delta_{ll'} \sim \epsilon_{l'}-\epsilon_l$,  we evaluate them within the FGR,
\begin{equation}
\Gamma_{ll'}= \mathrm{Re}  \int_0^\infty dt \mathrm{e}^{-i \Delta_{ll'} t} G_{ll'}(t), \
G_{ll'}(t) = 2 \langle \tilde H'_{l'l}(t)  \tilde H'_{ll'} \rangle, \label{gamma}
\end{equation}
where averaging is over the (boson) equilibrium at $T>0$.   
We simplify  $G_{ll'}(t)$ by neglecting  in Eqs.~(\ref{hbll}),(\ref{gamma}) cross-terms between $a_q$  and multi-boson 
$A_{ll'}$, 
\begin{eqnarray}
G_{ll'}(t) &=& 2 \sum_q |\eta_{ll' q}|^2 g_q(t) R_{ll'}(t), ~ R_{ll'}(t)= \mathrm{e}^{[Q_{ll'}(t)-Q_{ll'}(0)]},\nonumber \\
Q_{ll'}(t) &=& \sum_k |\zeta_{ll'k}|^2  g_k(t), \\
g_q(t) &=& (\bar n_q +1 ) \mathrm{e}^{-i \omega_q t} + \bar n_q \mathrm{e}^{i \omega_q t} , \nonumber
 \label{gammall}
\end{eqnarray}   
with  $\zeta_{ll'k}=\zeta_{l'k}-\zeta_{lk}$, and boson equilibrium occupation  $\bar n_q=1/[\mathrm{e}^{\omega_q/T}-1]$ .

\subsection{Simplified  transition rates} 
The above expressions for $\Gamma_{ll'}$  account for the details of the model and 
are rather complex. However, results may also be explained using more qualitative arguments. 
The essential ingredients within  the FGR are the conservation of energy and the overlaps between eigenstates 
$\phi_{li}, \phi_{l'i}$.  This  suggests  a  simplified form,
\begin{equation}
\Gamma^{\rm s}_{ll'} = B ~ \theta(\Omega- |\Delta_{ll'}|) M_{ll'},\label{toyhcb}
\end{equation}
where $M_{ll'}=\sum_i \phi^2_{li}\; \phi^2_{l'i}$ and  $\Omega \sim \omega_0 + 2t_b $ is the rigid cut--off 
for single-boson emission/absorption.
In order to account for  multi--boson processes  we can employ the saddle-point approximation 
(in analogy to  Ref. \onlinecite{lenarcic14}) to Eq. (\ref{gamma}) and $R_{ll'}(t)$, 
leading to  an exponential cut--off
\begin{equation}
\Gamma^{\rm m}_{ll'} = C   \mathrm{exp}\left[- \frac{a |{\Delta_{ll'}}|}{\omega_0}\right] M_{ll'}, \quad 
a = \ln \frac{\Delta_{ll'}}{e \nu \omega_0},  \label{toyph}
\end{equation}
with $\nu = (1/L) \sum_k |\zeta_{ll'k}|^2 \bar n_k \sim   2 \bar n g^2 \Omega^2$,
which we simplify further  taking $\omega_0/a \sim \Omega$.

\subsection{Rate equations}
 Within the FGR particle dynamics can be
described via the rate equations for occupations  $p_l(t)$,
\begin{equation}
d p_l/dt = \sum_{l'} ( \Gamma_{l'l} p_{l'} - \Gamma_{ll'} p_l). \label{req}   
\end{equation}   
In order to have a stationary solution, $p_l(t)=p_l^0$, rates $\Gamma_{ll'}$  should follow the detailed balance\cite{bouchaud89} 
$p_l^0/p_{l'}^0 = \Gamma_{l'l} / \Gamma_{ll'}$ for each pair $l,l'$. This is satisfied
within the form of Eq.~(\ref{gamma}) since to all orders in coupling $g$ one can show that 
\begin{equation}
\Gamma_{l'l} / \Gamma_{ll'} = (\bar n_q+1)/\bar n_q= \mathrm{e}^{\Delta_{ll'}/T},
\end{equation}
taking into account the energy conservation $\omega_q=\Delta_{ll'}$, see Eqs.~(\ref{gamma}),(\ref{gammall}). 
Then, at $T>0$  we obtain Boltzmann stationary state $p_l^0 = c ~\mathrm{e}^{-\epsilon_l/T}$ while 
Eqs.~(\ref{req}) can be symmetrized by introducing,
\begin{equation}
p_l(t)=\sqrt{p_l^0} \tilde p_l(t), \qquad \tilde \Gamma_{l'l} = \sqrt{p^0_{l'}/p^0_l} \Gamma_{l'l}.
\end{equation}
 The solution of  Eq.~(\ref{req}) can be generally represented in the form
$p_l(t)= \sum_m b_m p_{m l} \mathrm{e}^{- \Lambda_m t}$
with real and nonnegative $\Lambda_m$ due to the symmetric $\tilde \Gamma_{l'l}$, with $p_{m l}$ being 
corresponding eigenvectors, as well as with the lowest $\Lambda_1=0$. Further on, we study 
dynamical solutions for a particle starting from a single Anderson state, i.e.  $p_l(0)=\delta_{l,l_0}$. 

\subsection{Results} General characteristics of dynamical solutions can be extracted
from  $\Gamma_{l'l}$, in particular from the distribution of the total local transition rates
$\Gamma_l = \sum_{l'\neq l} \Gamma_{l'l}$.\cite{bouchaud89} In the following we calculate
the probability distribution ${\cal D}(\Gamma_l)$  for 
$\omega_0 =g=1, t_b = 0.4 $ on chain with $L=200-500$ sites. 
After finding numerically  SP states $\phi_{li}$, we evaluate all $\Gamma_{ll'}$ at chosen $T>0$, 
averaging also over $N_s  \gg 1$ realizations of disorder. 
Results presented  for integrated distribution
\begin{equation}
I(\Gamma_l) = \int_0^\Gamma  {\cal D}(\Gamma_l) d\Gamma_l
\end{equation}
are shown  (in log-log scale)  in Fig.~1a for $T=2$ and various disorders,
ranging from weak $W=2$ to strong $W=8$. For comparison, we display also corresponding 
results for simplified rates, Eq.~(\ref{toyph}),  for the same $W$ but adapted $C=6$, which, however,
 doesn't affect the structure of ${\cal D}(\Gamma_l)$. 
It is meaningful to interpret results in Fig.~1a  in terms of power-laws, i.e.,  
\begin{equation}
I(\Gamma_l) \propto  \Gamma_l^\alpha, \qquad {\cal D}(\Gamma_l) \propto  \Gamma_l^{\alpha-1}.
\end{equation} 
The corresponding distribution  ${\cal F}(\tau_l)$ for the local lifetimes $\tau_l=1/\Gamma_l$ can be obtained by 
comparing  $I(\Gamma_l) =\int_{1/\Gamma_l}^{\infty}  {\mathrm d} \tau_l  {\cal F}(\tau_l) $, leading to
${\cal F}(\tau_l) \propto \tau_l^{-(\alpha+1)}$.  Results for $I(\Gamma_l)$ 
will be further related to the straightforward calculation of the SP spread  
$\sigma^2(t) = \sum_l (l-l_0)^2 p_l(t)$ presented in Fig.~1c for the same parameters.

Different regimes in Fig.~1a can be analyzed in terms of the classical random-trap model.\cite{machta85,bouchaud89}
Normal diffusion is the solution for $\alpha > 1$ leading to a finite  average local lifetime 
\begin{equation}
\bar{\tau}  = \int_{\tau_{\rm min}}^\infty d\tau_l \; \tau_l {\cal F}(\tau_l) < \infty,
\end{equation}
and $\gamma=1$, i.e., the spread $\sigma^2(t) \propto  D t$ with the diffusion constant $D \propto  1/\bar{\tau}$.   
In Figs.~1a and  1c this is the  case for $W<W^* \sim 4$, although quite long times $t \gg 100$ are needed 
to confirm $\gamma \sim 1$. 

Here, we  are mostly interested  in the anomalous subdiffusive dynamics, which is the case for  $0<\alpha <1$. 
If valid in the whole regime  $\Gamma_l \to 0$ (or equivalently for  $\tau_l \to \infty$ ) this would imply infinite  $\bar{\tau}$.  
We note that in Fig.~1a,  $\alpha <1$ appears for $W>W^*$ 
within large span of $\Gamma_l > \Gamma^*$. Threshold rate $\Gamma^*$ strongly decreases with $W$  and  for $W>8$  
it is below the numerical accuracy of the present calculations.
Nevertheless, for $\Gamma_l < \Gamma^*$ we observe $\alpha > 1$.  Therefore $\bar{\tau}$ is huge but finite, suggesting that 
the subdiffusion is a transient  phenomenon and the dynamics should eventually become normal diffusive. 
In Fig.~1c it is visible that subdiffusive $\gamma <1$ indeed appears for $W>W^*$. 
Still, the normal diffusion may be visible only for  long chains $L > 1/I(\Gamma^*) $  and very long times 
$t  \gg  \bar{\tau} > 1/\Gamma^*$.

\begin{figure}
\centering
\includegraphics[width=\columnwidth]{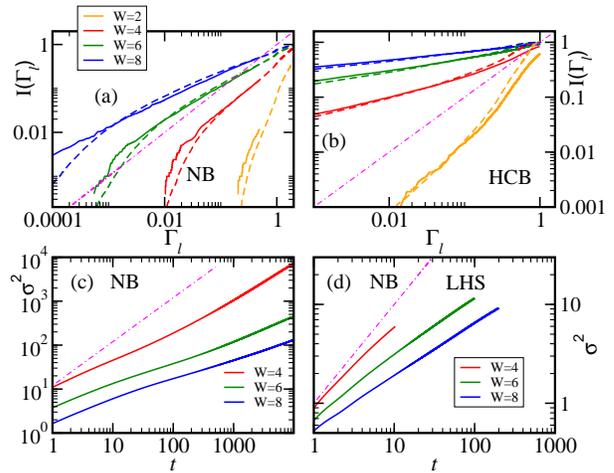}
\caption{a) and b): Integrated distributions of  local rates $I(\Gamma_l)$ at different disorders $W$,  for 
$\omega_0=g=1, t_b=0.4$ with full (continuous lines) and simplified (dashed lines) rates:
a) NB at $T=2$ and b)  HCB at $T \rightarrow \infty$.   c) Time evolution of the SP spread  $\sigma^2(t)$ for NB 
at the same parameters,  calculated via rate equations, d) evolution of the full many-body system using the LHS method. 
Dot-dashed lines show the diffusion thresholds.}   
\label{fig1}
\end{figure} 

In order to test the feasibility of the FGR and rate-equation approach, we study also directly the evolution of the coupled
particle-boson many-body system. The time evolution of the whole system is performed in analogy to previous works 
\cite{bonca17, lemut17,bonca18} by using  limited Hilbert-space (LHS) method, where we start with a particle  
at single site and  $N_b >0$ initial bosons in a system of finite effective size $L \sim 16$. Results evaluated at disorders $W=4 - 8$ are 
presented in  Fig.~1d and are qualitatively in agreement with results in Fig.~1c taking into account that  LHS allows only for restricted 
sizes and consequently limited $t$.  In particular, LHS results confirm the (transient) subdiffusive dynamics
with $\gamma <1$  for $W=6,8$, while diffusive regime cannot be reached due to small $L$ as well as too short $t \ll \bar{\tau}$. 

\section{Hard-core bosons} Due to reduced Hilbert space, the model with HCB offers the advantage for 
full many-body simulations.\cite{bonca17,bonca18} Moreover, the connection of HCB to 
spin systems in 1D allows  closer relation with the disordered Hubbard model\cite{mondaini15,prelovsek16,kozarzewski18} 
and the disordered Heisenberg model.\cite{oganesyan07,luitz16,prelovsek116}
Using the standard relation of HCB with $S=1/2$ local spin operators,
we can follow previous procedure and eliminate the diagonal $l'=l$ term via local
transformation $U_l=\prod_i U_{li}$,
\begin{equation}
U_{li}=  \cos(\mu_{li}) + 2 i \sin(\mu_{li}) S_i^y, \quad \mathrm{tg}(2\mu_{li}) =  -2g \frac{\eta_{li}}{\omega_0},
\end{equation}
leading instead of Eq.~(\ref{hbll}) to
\begin{eqnarray}
\tilde H^h_{l l'}&=& 2 \sum_i \eta_{l l' i}  U_{ll'i} \tilde S_{i;ll'}^x, \quad U_{ll'i}= \prod_{j \neq i} U_{l'j} U^*_{lj}, 
\nonumber \\
\tilde S_{i;ll'}^x &=& U_{l'i} S_i^x U^*_{li} = \cos (\mu_{ll'i}) S_i^x + \sin (\mu_{ll'i}) S_i^z, \label{hcb} 
\end{eqnarray}
where  $\mu_{ll'i}= \mu_{l'i}+\mu_{li}$. In spite of formal similarity to NB, Eq.~(\ref{hbll}), 
there is essential difference, that in Eq.~(\ref{hcb}) multi-boson processes are strongly reduced, i.e.,
$\tilde H^h_{l l'}$ allows at most a single boson creation/annihilation per site $i$ of state $\phi_{li}$.
In case of strong disorder with short localization length $\xi \sim 1$ this eliminates to large extent multi-boson processes 
within FGR, hence we simplify Eq.~(\ref{gamma}) by replacing $U_{ll'i} \sim 1$. 
Within the same spirit we assume  in Eq.~(\ref{hcb}) $|\mu_{ll'i}| \ll 1$  and $\tilde S_{i;ll'}^x \sim S_i^x$.  
Standard  transformation of 1D spin operators to fermions then yields,
\begin{equation}
G_{ll'}(t)  = \frac{2}{L} \sum_q |\tilde \eta_{ll'q}|^2  [f_q \mathrm{e}^{-i \omega_q t} + (1-f_q) \mathrm{e}^{i \omega_q t} ],
\label{gamhcb}
\end{equation}   
with Fermi-Dirac distribution $f_q=1/[\mathrm{e}^{\omega_q/T} +1]$.

Results for HCB can be now evaluated and analyzed in analogy to NB case. One advantage is
that $T$ is less relevant for HCB and we can directly take $T \to \infty$ (as mostly considered in MBL studies)
by inserting $f_q=0.5$ in Eq.~(\ref{gamhcb}).  From the distribution of $\Gamma_l$ presented in Fig.~1b,
the difference to NB is obvious. Namely, due to suppressed multi-boson processes, the distribution of ${\cal D}(\Gamma_l)$ 
can be singular with $\alpha <1$  in the whole range of $\Gamma_l > 0$, at least for large enough $W > W^* \sim 3$ (for considered
parameters). This emerges also for the simplified  rates, Eq.~(\ref{toyhcb}) with the prefactor $B=1.5$, where the choice of $B$ sets 
only the time-scale.
At the same time, the spread as shown in Fig.~2a reveals consistently only subdiffusion with 
$\sigma^2 \propto t^\gamma, ~ \gamma<1$, with no crossover to normal diffusion, in contrast to Fig.~1c.  
This confirms a nontrivial phenomenon, that  coupling to HCB leads to thermalization
 (here at $T \to \infty$), but still not to normal diffusion 
at long times. In other words, in the case of HCB  the dynamics  can remain subdiffusive in 1D. 

In Fig.~2b we present the corresponding probability profiles $\bar p_{l}(t)$ (with the reference starting site $l_0=0$)
averaged over all initial sites, at fixed time $t=50$ and different $W$. It is characteristic that the profiles 
deviate from a normal Gaussian and  become almost exponential $\bar p_l \propto \exp( - \lambda(t) |l|)$ for strong disorder.
Moreover, $\bar p_{l}(t)$ reveals at all $W, t$ an evident deep at $l=1$, due to nearest--neighbor states
being too far in energy, $|\epsilon_{l+1} - \epsilon_{l}| > 2 t_h > \Omega$,\cite{mott68} to contribute to $\Gamma_{l,l+1}$.

\begin{figure}
\centering
\includegraphics[width=\columnwidth]{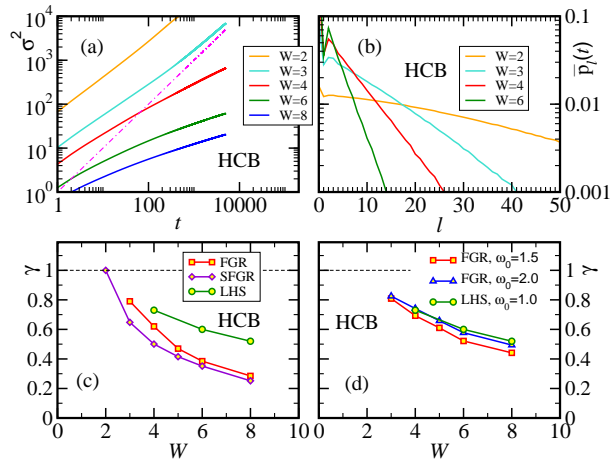}
\caption{ a) Spread $\sigma^2(t)$, evaluated for HCB with parameters as in Fig.~1b,  b) corresponding averaged SP profile $\bar p_l(t)$, 
evaluated for fixed $t=50$, and  for different $W=2-6$,  c) (sub)diffusion exponents $\gamma$ vs. disorder $W$ as evaluated from FGR, simplified rates (SFGR),
and via full simulation using LHS, respectively. d) $\gamma$ from  FGR approach for different 
$\omega_0=1.5, 2.0$ and shorter $t<100$ compared with LHS results for $\omega_0=1$.}
\label{fig2}
\end{figure} 
Finally, Fig.~2c shows the comparison between exponents $\gamma$, as obtained for HCB case from different 
methods again for $\omega_0=g=1, t_b=0.4$ but varying disorder $W$: a)  numerical simulations via  LHS followed 
to distances $\sigma \sim 5$;  b) the spread $\sigma^2(t)$ emerging from FGR and rate equations for size $L \sim 200$;
and c) simplified equation (\ref{toyhcb}) via extracting $\alpha$  from the tails of  
$I(\Gamma_l) \propto \Gamma_l^{\alpha}$ for $10^{-4}< \Gamma_l<10^{-2}$
and  taking into account the relation   for  classical random-walking, \cite{machta85,bouchaud89}
\begin{equation}
\gamma=2\alpha/(1+\alpha).
\end{equation}
We can notice that the full and simplified FGR results do agree well, while $\gamma$ from the full many-body time 
evolution is still significantly larger. The quantitative disagreement partly emerges from 
restricting Eq.~(\ref{gamhcb}) to stritctly single-boson processes, whereby two-boson processes might also contribute.
This can be effectively simulated by increasing $\omega_0$. We therefore present in Fig~3d the FGR results also for  
$\omega_0=1.5, 2.0$ (and $t_b=0.6, 0.8$, respectively) which reveal  better match with numerics.

\section{Quasi-periodic potential.} 
\begin{figure}
\centering
\includegraphics[width=\columnwidth]{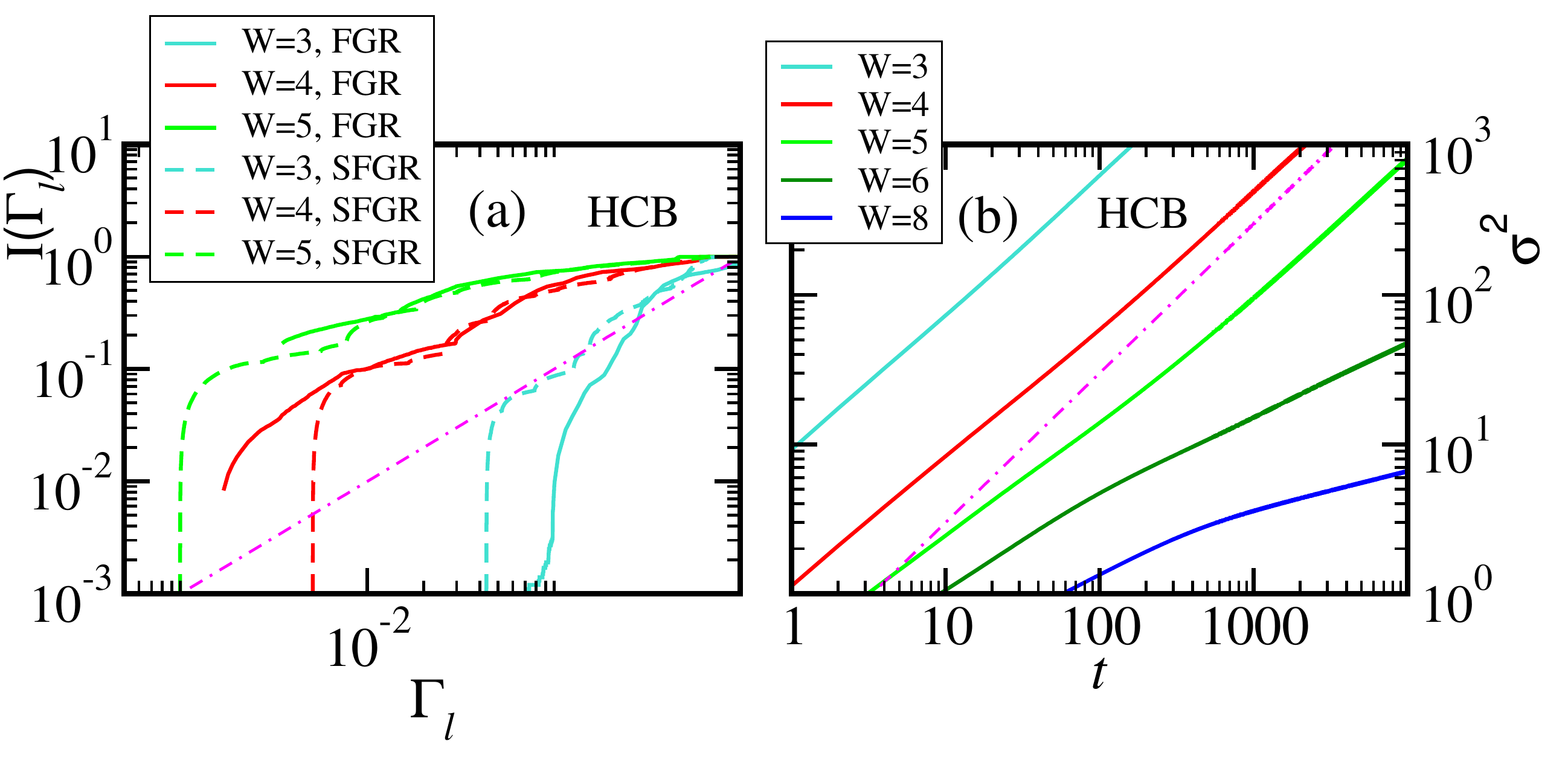}
\caption{a) Distribution $I(\Gamma_l)$ and b)  spread $\sigma^2(t)$ for HCB with the same parameters as in Fig.~1b but
for quasi-periodic potential. }
\label{fig3}
\end{figure}

In order to elucidate further  the subdiffusion in the case of HCB, 
we compare results with the model with quasiperiodic potential, as actually realized in cold-atom 
experiments  on optical lattices.\cite{schreiber15,bordia16,luschen17} We choose it in the (Aubry-Andree)  form
$h_i=\sqrt{2/3} W \cos(2 \pi q_0 i+\psi_0)$ which has the same standard deviation as the random one, where 
$q_0=(1+\sqrt{5})/2$ is a golden mean and $\psi_0$ an arbitrary phase. We note in Fig.~3a that the distribution 
$I(\Gamma_l)$ is qualitatively different from a random potential in Fig.~1b. The difference emerges from 
correlated energies allowing for resonance contributions to $\Gamma_l$.  This indicates that  for quasi-periodic potential  
the long-time dynamics would be always  diffusive\cite{znidaric18} even for large $W$ , although from $\sigma^2(t)$ in Fig.~3b
this is expected to emerge for, e.g., $W=6,8$ only for extremely long $t\gg \bar{\tau} \gg 10^2$.
It is quite clear  that the same conclusion holds true also for a particle in the quasiperiodic potential 
coupled to noninteracting bosons. In the latter case, the multi-boson  processes  which are suppressed  for HCB, additionally contribute 
to the diffusive transport in the long--time regime.  Results  (not presented here) for quasiperiodic potential and  NB are qualitatively very similar to results shown in Fig. \ref{fig3} for HCB.

\section{Conclusions} The discussed model of SP moving in a 1D random potential is 
a prototype problem of quantum propagation in a disordered medium due to  coupling to other (bosonic) 
dispersive degrees of freedom. We show that results obtained via the FGR represent an important
simplification and insight, still they are nontrivial and appear to well (at least qualitatively) describe the whole
many-body dynamics, as simulated numerically. First, due to the coupling to bosonic subsystem, the particle 
evolution is ergodic, approaching thermal
equilibrium for $ t \to \infty$. Still, a diffusive dynamics is not a rule. For noninteracting 
bosons it can appear only after transient subdiffusive spread $\sigma(t) \propto t^\gamma$ with
$\gamma <1$, where time span of the latter  regime strongly depends on the disorder $W$ and bosonic temperature $T$. 
Moreover, in the case of HCB our analysis and numerical results indicate  that the subdiffusion persists at longest 
times, whereby the difference emerges due to multi-boson  processes which are allowed for noninteracting bosons but  are strongly suppressed for HCB.
Still, beyond the energy conservation the character of SP wavefunctions are also crucial, as evident from the
result that in a quasi-periodic potential the subdiffusion is only a  transient phenomenon\cite{znidaric18} also for HCB.
 
 One cannot exclude that within a more accurate treatment of the multi-boson processes, the 
normal diffusion eventually  sets on also in the HCB model. However, the crossover to normal diffusion will then happen at the time-scales which are much longer than for NB and, most probably, would be irrelevant for experiments. 

\acknowledgments  
 P.P. and J.B. acknowledge the support by the program P1-0044 of the
Slovenian Research Agency. M.M. is supported by the National Science Centre, Poland via project 2016/23/B/ST3/00647.

\bibliography{ref_mbl}

%%%%%%%%%% Merge with supplemental materials %%%%%%%%%%
%%%%%%%%%% Merge with supplemental materials %%%%%%%%%%
\end{document}